\documentclass{eptcs}
\providecommand{\event}{TFPIE 2018} 
\usepackage{breakurl}             
\usepackage[noxcolor]{pstricks}
\usepackage{pst-node}
\usepackage{graphicx}
\usepackage{subfigure}
\usepackage{stfloats}
\usepackage{multido}
\usepackage{setspace}
\usepackage{textcomp}
\usepackage{amssymb}
\usepackage{alltt}
\usepackage{pgfplots}
\pgfplotsset{compat=newest}
\usepackage{pgfplotstable}
\usepackage{pgf-pie}
\usepackage{textcomp}

\newcommand{\fsm}{\textsf{FSM} \space}
\newcommand{\fsmn}{\textsf{FSM}}
\newcommand{\racket}{\textsf{Racket} \space}

\title{FSM Error Messages}

\author{Marco T. Moraz\'an
\institute{Seton Hall University}
\email{morazanm@shu.edu}
\and
\and
Josephine A. Des Rosiers
\institute{Seton Hall University}
\email{desrosjo@shu.edu}
}
\def\titlerunning{FSM Error Messages}
\def\authorrunning{Moraz\'an and Des Rosiers}

\begin{document}
\maketitle

\begin{abstract}
Computer Science students, in general, find Automata Theory difficult and mostly unrelated to their area of study. To mitigate these perceptions, \fsmn, a library to program state machines and grammars, was developed to bring programming to the Automata Theory classroom. The results of the library's maiden voyage at Seton Hall University had a positive impact on students, but the students found the library difficult to use due to the error messages generated. These messages were generated by the host language meaning that students needed to be familiar with the library's implementation to make sense of them. This article presents the design of and results obtained from using an error-messaging system tailor-made for \fsmn. The effectiveness of the library was measured by both a control group study and a survey. The results strongly suggest that the error-messaging system has had a positive impact on students' attitude towards automata theory, towards programming in \fsmn, and towards \fsm error messages. The consequence has been a marked improvement on students' ability to implement algorithms developed as part of constructive proofs by making the debugging of \fsm programs easier.
\end{abstract}

\maketitle

\section{Introduction}
Programming-oriented students usually find automata theory, grammars, constructive proofs, computability, and decidability challenging and in many cases overwhelming. The existence of this perception is not completely absurd since students are asked to design and prove correct machines/grammars/algorithms without being able to experiment nor get immediate feedback, as they do in any programming course. Both of these are essential in a learning context and need to be timely (the sooner the better) \cite{Race}. In fact, designing programs without being able to experiment and to get immediate feedback in the form of \emph{understandable} error messages goes against the grain of what students learn in programming courses. Given that formal automata textbooks (e.g., \cite{Lewis,Martin,Sipser}) rarely offer any software infrastructure for students to experiment with their ideas and designs, it comes as no surprise that students dislike the material. More importantly, however, this has an impact on the quality of the constructive proofs they develop and, therefore, the grade they receive. Students may have a general idea for an algorithm, but are fuzzy about the details. Having them code the machines/grammars/algorithms they develop (as part of a constructive proof) forces them to think about the details and improves the quality of their proofs resulting in better grades and less frustration.

Given that it is reasonable for Computer Science students to be able to experiment with their designs, the \textsf{FSM} (\textbf{F}unctional \textbf{S}tate \textbf{M}achines) library was developed \cite{fsm}. This library (implemented in \textsf{Racket}) allows students to implement state-machines (e.g., finite-state automata, pushdown automata, and Turing machines) and grammars (e.g., regular, context-free, and context-sensitive). In addition, the library implements many common constructors (e.g., union, concatenation, and complement) derived from constructive proofs developed in class which students can use as part of their designs. The library also provides useful functions, like renaming the states of a machine and renaming the nonterminals of a grammar, to allow students to focus on the algorithms they develop as part of their constructive proofs instead of focusing on the auxiliary functions they may need. Finally, \textsf{FSM} provides random testing facilities for both machines and grammars which students (and instructors) can use to validate (not verify) a machine or grammar. The expectation is three-fold. First, we expect CS students to develop a more favorable attitude towards the material as they are able to implement and test their algorithms. To the best knowledge of the authors, \textsf{FSM} is the only library of its kind that provides students with an appropriate level of abstraction to achieve this. That is, students do not need to, for example, wrestle with how to implement nondeterminism. Second, given that students can debug their algorithms before submitting them for grading the expectation is that more solutions to problems will be correct. Third, grading is simplified for the instructor by being able to test the, sometimes convoluted, solutions students develop.

The maiden voyage of \textsf{FSM} at Seton Hall University (\textsf{SHU}) took place in the Spring semester of 2016 and was taught by the first author. The results, at best, did not fully meet the expectations. Although students seemed to develop a more favorable attitude towards Automata Theory, there was still a significant number of students that simply did not relate to the material. Grades were also, generally speaking, higher thanks to the library, but mostly for students that made an effort to see the instructor outside of class or to ask questions during class. The overwhelming number of questions were about how to use \textsf{FSM}. Finally, grading was not significantly simplified. In many instances, it was very difficult to explain why the constructors for state machines and for grammars submitted by students were buggy.

The lukewarm results for all three expectations were rooted in the lack of a proper error-messaging system for \textsf{FSM}. In general, the errors that students (and the instructor) received came from the \textsf{Racket} error-messaging system and were rarely useful in diagnosing the error at the \textsf{FSM} level. For example, consider applying a machine \textsf{M} to a word \textsf{w}. If \textsf{M} is in state \textsf{A} and the next element of \textsf{w} to be processed is \textsf{b}, \fsm filters \textsf{M}'s rules to find the ones that apply to such a configuration. The rules are then applied to create a list of next possible configurations (recall that \textsf{M} may be nondeterministic). A student, however, has mistyped the state \textsf{A} as \textsf{a} in the rules. Thus, the filtered list of rules is empty. This leads to a \textsf{Racket} error stating that \textsf{car} cannot be applied to the empty list when \fsm attempts to create the list of next possible configurations. To understand such an error, the programmer has to be intimately familiar with the implementation of \textsf{FSM}, which is an unreasonable expectation for students in the course (just like being familiar with the implementation of \textsf{Racket} is also unreasonable). A more useful error ought to be produced by the library when the constructor for the machine was called with rules that have typos.

This article presents the design of an error-messaging system for \fsm and the impact it has had on students. The focus is on generating error messages that are meaningful to students when they improperly use a constructor. Section \ref{rw} reviews related work on designing error messages. Section \ref{ov} provides a brief overview of \fsmn. Section \ref{pe} presents examples of code developed by students and the \textsf{Racket} error messages they originally received. Section \ref{fsmerrs} presents the design of \fsm error messages. Section \ref{fb} presents feedback received from the students. This section is divided in two subsections. Section \ref{cg} presents the results obtained from a control-group study performed by asking students to debug the solutions to three problems. The control group worked with a version of \fsm that did not have a specialized error-messaging system and the focus group worked with one that did. Section \ref{ss} presents results obtained from surveying the students. Finally, Section \ref{fw} briefly presents some conclusions and directions for future work.

\section{Related Work}
\label{rw}
There are two primary sources in the literature that influenced the design of the \fsm error-messaging system. The first is the work done by developers of programming languages for students. The second is the work done by the human-computer interaction (\textsf{HCI}) community on compiler error messages. Both communities have developed principles to guide implementors of error-messaging systems. These communities coincide in stating that experienced programmers can often see mistakes that students do not and in stating that error messages need to be understandable \cite{Crestani,Hsia,Munson,Schiliep,Traver}. These two conclusions come as no surprise, but how to make error messages understandable, especially to students, is yet unresolved.

A fundamental principle is that how error messages are phrased is important \cite{McIver,Traver}. For students, this means that error messages must not be filled with technical jargon that is unfamiliar to them. In fact, error messages must be recognized as the primary feedback mechanism for students \cite{Marceau,Mind,Munson}. This is evident, because when solving homework problems, students spend more time wrestling with error messages than they do asking an instructor what an error message means. To assist novice programmers it is important for error messages to be expressed using the same vocabulary that is employed in class by the instructor \cite{Marceau}. In this manner, error messages may become meaningful to students by eliminating unfamiliar technical jargon. The work presented in this article is very much inspired by this principle. The vocabulary used by the instructor in class is the same vocabulary that students see in an error message. Furthermore, the instructor on occasion purposely presents buggy code to the students in order to demonstrate what the error messages mean and to explicitly show students that the vocabulary in the error messages is the vocabulary used in class. The instructor makes a conscious effort to use terms that appear in error messages (e.g., states, alphabets, and rules) instead of using technical jargon that explains why in the \fsm implementation incorrect inputs to a constructor lead to errors when applying machines and grammars. The set of terms used are unified over the textbook, the error messages, and the lectures.

Error messages must also be carefully designed to prevent students from taking random actions to correct a bug \cite{Traver}. To this end, it is important for error messages to provide guidance as to why the problem exists so that the student/programmer can make progress in resolving it \cite{Marceau,Mind,Traver}. Poor error messages (especially those that prescribe solutions) lead to actions that do not bring programmers, especially novices, closer to a solution. For example, \fsm can be used to construct a deterministic finite-state machine where the set of states is \textsf{\textquotesingle(X Y Z)} and the alphabet is \textsf{\textquotesingle(a b)}. A programmer may then mistakenly define a transition rule to move from state \textsf{X} to state \textsf{Z} on \textsf{a} as \textsf{\textquotesingle(x a Z)}. An error message generated when the machine is applied to some input stating that the list of rules is empty is a poor message. It leads students to edit the design of their machines. A more useful error message is to have the constructor for deterministic finite-state machines throw an error stating that:
\emph{in the rule \textsf{\textquotesingle(x a Z)}, \textsf{x} is not a valid state}. In fact an error message should not propose a solution \cite{Marceau}. For the example above, there are at least three possible corrective actions: add \textsf{x} to the set of states, change \textsf{x} to \textsf{X} in the rule, or change \textsf{x} for some other state in the rule. There is no way for the error-messaging system to know which corrective action should be taken. This principle is another of the guiding pillars of the \fsm error-messaging system.

There is some debate on whether longer or shorter error messages are better. Some argue that longer error messages are not better \cite{Nienaltowski}. Intuitively, this can be explained by students not willing to read error messages that they do not understand (i.e., the used vocabulary is foreign to them). If this is the case, then it is difficult to discern if the length of a message actually matters. On the other hand, short error messages may be too brief and not provide enough information to fix a bug. The problem is compounded when you need dependent types to describe the inputs to a function. For example, the set of rules given as input to the constructor for a deterministic finite-state machine depends on the given set of states and the given alphabet. If the set of rules is invalid, should an error message be short and simply state that the set of rules is invalid or should the error message be longer and state which and why rules are invalid? In \fsmn, we have chosen to have longer error messages. This decision is based solely on informal discussions with students that state that if error messages were actually useful, then they would be willing to read them.

\begin{figure}
\begin{alltt}
     (define sol1-dfa (make-dfa \textquotesingle(q0 q1 q2 ds) ;; the set of states
                                \textquotesingle(a b)         ;; the input alpahbet
                                \textquotesingle{q0}            ;; the starting state
                                \textquotesingle(q1)          ;; list of final states
                                \textquotesingle((q0 a q1)    ;; list of transition rules
                                  (q0 b ds)
                                  (q1 a q1)
                                  (q1 b q2)
                                  (q2 a q1)
                                  (q2 b q2)
                                  (ds a ds)
                                  (ds b ds))))

     (check-expect (sm-apply sol1-dfa \textquotesingle{(a)}) \textquotesingle{accept})
     (check-expect (sm-apply sol1-dfa \textquotesingle{(a b b b a)}) \textquotesingle{accept})
     (check-expect (sm-apply sol1-dfa \textquotesingle{(a b a a b b)}) \textquotesingle{reject})
\end{alltt}
\caption{\fsm implementation of a deterministic finite-state machine to recognize a(a $\cup$ b)$^*$a.}
\label{dfa1}
\end{figure}

\section{Overview of \fsm}
\label{ov}
The \fsm library presents the user with a generic interface to construct and manipulate state machines and grammars. Constructors are divided into two categories: primitive constructors and transformers. Primitive constructors build a state machine or a grammar from a formal description provided by the programmer. Transformers build a state machine or a grammar from existing machines or grammars exploiting algorithms obtained from constructive proofs. Observers are divided into three categories: accessors, applicators, and testers. Accessors return a specified component used to build a grammar or a state machine. Applicators apply a given machine or grammar to a word. Testers allow for machines and grammars to be tested with words provided by the programmer or with randomly generated words by the software. The latter two provide students with immediate feedback on the validity (not the verification) of their designs and implementations. In this section, we restrict ourselves to a couple of examples to give the reader a flavor of how \fsm primitive constructors are used. For a full description of \fsmn, the reader is referred to a previously published article \cite{fsm}.

Let $\Sigma = \{a, b\}$ be an input alphabet. Consider the problem of recognizing the regular language:
\begin{quote}
$L = \{$$w | w \in \Sigma^* \wedge$ \emph{w starts and ends with an a}$\}$.
\end{quote}
In \fsmn, the finite-state machine is constructed as displayed in Figure \ref{dfa1}. We can describe each state with an invariant. Informally, \textsf{q0} is the starting state representing that nothing has been read. The state \textsf{q1} represents that the input read so far starts and ends with an \textsf{a}. The state \textsf{q2} represents that the input read so far starts with an \textsf{a} and does not end with \textsf{a}. Finally, the dead state, \textsf{ds}, represents that the input does not start with an \textsf{a}. Based on these invariants, it is clear that the only final (or accepting) state can be \textsf{q1}. Each transition specifies a from-state, the input read on the tape, and a destination-state. For example, \textsf{\textquotesingle(q0 a q1)} says that from state \textsf{q0} upon reading an \textsf{a} the machine moves to state \textsf{q1}.

The programmer can test the machines with, for example, 5 randomly generated inputs as follows:
\begin{alltt}
     > (sm-test sol1-dfa 5)
     \textquotesingle(((b a a a a b b) reject)
       ((b a b b b a) reject)
       ((a a b a a a b a) accept)
       ((a b a a a b) reject)
       ((a a) accept))
     >
\end{alltt}

As a second example, consider implementing a context-free grammar for the set of palindromes over the same $\Sigma$ above. In \fsmn, the context-free grammar is constructed as follows:
\begin{alltt}
     (define palindrome (make-cfg \textquotesingle(S)    ;; the set of nonterminal symbols
                                  \textquotesingle(a b)  ;; the alphabet
                                  \`{}((S ,ARROW ,EMP) ;; the set of rules
                                    (S ,ARROW a)
                                    (S ,ARROW b)
                                    (S ,ARROW aSa)
                                    (S ,ARROW bSb))
                                   \textquotesingle{S}))   ;; the starting nonterminal
\end{alltt}
The constants \textsf{ARROW} and \textsf{EMP} are defined by \fsm to denote $\rightarrow$ and, the empty string, $\epsilon$, respectively.The programmer can now test the grammar with 3 random words as follows:
\begin{alltt}
     > (grammar-test palindrome 3)
     \textquotesingle(((b b a a a) "(b b a a a) is not in L(G)")
        ((b a a b) (S -> bSb -> baSab -> baab))
        ((b b b) (S -> bSb -> bbb))
     >
\end{alltt}
As the reader can see, if a word is in the language the tester provides its derivation. Otherwise, the tester explicitly states the word is not in the language of the grammar.

\begin{figure}
\begin{alltt}
          (define has-bb (make-tm \textquotesingle(S A B Y N)         ;; the set of states
                                  \textquotesingle{S}                   ;; the starting state
                                  \textquotesingle(a b)               ;; the alphabet
                                  \textquotesingle(Y N)               ;; the halting states
                                  \`{}(((S b) (A ,RIGHT)) ;; the transition rules
                                    ((S a) (S ,RIGHT))
                                    ((S ,BLANK) (N ,BLANK))
                                    ((A b) (B ,RIGHT))
                                    ((A a) (S ,RIGHT))
                                    ((A ,BLANK) (N ,BLANK))
                                    ((B a) (B ,RIGHT))
                                    ((B b) (B ,RIGHT))
                                    ((B ,BLANK) (Y ,BLANK)))
                                  \textquotesingle{Y}))                 ;; the accepting state

          (check-expect (sm-apply has-bb \textquotesingle{(a b a a b a b b a a)}) \textquotesingle{accept})
          (check-expect (sm-apply has-bb \textquotesingle{(a b a a b a a a)}) \textquotesingle{reject})
\end{alltt}
\caption{Buggy Turing machine that decides the language L = \{w $|$ w has two consecutive b's\}}.
\label{tmaa}
\end{figure}

\section{Student Programming Examples with \racket Errors}
\label{pe}
This section presents and discusses the error messages obtained with three sample student programs that contain bugs. The error messages in this section are those obtained before the \fsm error-messaging system was developed. You do not need to be familiar with \racket as the error messages are explained with a mostly jargon-free description.

Figure \ref{tmaa} displays the implementation of a Turing machine that decides the language of all strings in \textsf{(a, b)$^*$} that have two consecutive \textsf{b}. The student has correctly designed the transition function that moves from the starting state, \textsf{S}, on a \textsf{b} to a state, \textsf{A}, meaning that a \textsf{b} has been read. From \textsf{A} on a \textsf{b} the machine moves to a state, \textsf{B}, meaning that two consecutive \textsf{b}'s have been read. From \textsf{A} on an \textsf{a} the machine moves to \textsf{S} meaning that no \textsf{b}'s for \textsf{bb} have been read. Both states, \textsf{S} and \textsf{A}, on a blank correctly move to state \textsf{N} representing the reject state as \textsf{bb} is not contained in the input word. From state \textsf{B}, the machine reads the rest of the word and upon reading a blank moves to the accepting state, \textsf{Y}, and halts. It is unimportant that reading the rest of the input is unnecessary and a more ``efficient'' Turing machine would directly move from \textsf{A} on a \textsf{b} to \textsf{Y} and halt (eliminating the need for \textsf{B}). The student has even included a pair of tests to demonstrate their understanding of how the machine ought to behave.

The problem with this solution is that 4 of the arguments to the constructor, \textsf{make-tm}, are in the incorrect position. The correct order for the arguments is: the list of states, the alphabet, the transition rules, the starting state, the list of halting states, and the accepting state. When the tests are executed, the following error message is obtained:
\begin{alltt}
     caar: contract violation
     expected: (cons/c pair? any/c)
     given: \textquotesingle{a}
\end{alltt}
Clearly, this error is not useful in determining that the arguments to the constructor are incorrect unless you are familiar with the implementation of \fsmn. \fsm is reporting (via the \racket error-messaging system) an error that occurs while attempting to parse the argument given for the list of transition rules. There is no way for a programmer unfamiliar with the implementation of \fsm to know how to correct this unless they are told ``what the error really means." It is noteworthy that others have also found that reporting parsing errors is problematic for novices \cite{Mind}.

Figure \ref{pali} displays a student's proposed context-free grammar for the language of palindromes (discussed in Section \ref{ov}). In addition to the context-free grammar, the student includes an example of a word that ought to be successfully derived using an \fsmn-defined tester. There are four errors in this grammar:
\begin{description}
  \item[Capitalization] \fsm convention is that nonterminals need to be capitalized and terminals need to be lower case. In this example, the list of nonterminals needs to be capitalized. In the third rule, the \textsf{B} needs to be lower case and the \textsf{S} needs to be capitalized.
  \item[Unknown symbols] The list of rules contains the undefined symbol \textquotesingle{EMP}. Presumably, the student meant to use \textsf{quasiquote}, instead of \textsf{quote}, to define this list. The symbol \textquotesingle{ARROW} is also unknown, but given that the \fsm rule parser ignores the middle component of a rule it is something that in practice has no effect.
  \item[Starting symbol] The argument for the starting symbol is a list instead of a symbol.
  \item[Incorrect Language] The grammar only generates palindromes of even length.
\end{description}
The initial error reported is:
\begin{alltt}
     symbol->string: contract violation
     expected: symbol?
     given: \textquotesingle{(S)}
\end{alltt}
Once again, we see \fsm reporting a parsing error. Although the error reported is concerning the \racket function \textsf{symbol-\tt{>}string}, luckily this time the error indicates a problem with the given argument for the starting symbol and the student changes the list to the symbol \textsf{\textquotesingle{S}}.

\begin{figure}
\begin{alltt}
     (define palindrome (make-cfg \textquotesingle(s b)            ;; the set of nonterminals
                                  \textquotesingle(a b)            ;; the alphabet
                                  \textquotesingle((S ARROW EMP)   ;; the list of rules
                                    (S ARROW aSa)
                                    (S ARROW Bsb))
                                  \textquotesingle(S)))            ;; the starting nonterminal

     (grammar-derive palindrome \textquotesingle(a b b a))
\end{alltt}
\caption{Student's proposed context-free grammar for $L = \{w \in (a, b)^* | w \ is \ a \ palindrome\}$.}
\label{pali}
\end{figure}

After making this correction, no errors are reported. The student's example has the following result:
\begin{alltt}
     "(a b b a) is not in L(G)"
\end{alltt}
Why is the student confused by this? The student is expecting some error to help solve the problem. In this case, \fsm ought to provide more help given that the constructor is incorrectly used. For example, one error is that \textquotesingle{S} is not in the set of nonterminals, but appears as a nonterminal in the set of rules.

\begin{figure}
\begin{alltt}
;remove-unreachable : dfa --> dfa
;Purpose: To remove the unreachable states from the given dfa
(define (remove-unreachable m1)
  (letrec
      (;reachable-states : (listof Rules)(listof States) (listof States)
       ;                   --> (listof States)
       ;Purpose: To determine the reachable states from the given list of
       ;         states and rules
       (reachable-sts (lambda (lor tovisit reachable)
          ; Accumulator invariants
          ;   tovisit: reachable states whose successors have not been found
          ;   reachable: reachable states whose succesors have been found
          ...))

       (new-states (reachable-sts (sm-getrules m1)
                                  (list (sm-getstart m1))
                                  \textquotesingle())))
    (make-dfa new-states
              (sm-getalphabet m1)
              (sm-getstart m1)
              (sm-getfinals m1)
              (sm-getrules m1))))

(define dfa-test1 (make-dfa \textquotesingle(q0 q1 q2 q3 q4 ds)
                            \textquotesingle(a b)
                            \textquotesingle{q0}
                            \textquotesingle(q1 q3)
                            \textquotesingle((q0 a q1) (q0 b ds) (q1 a q1)
                              (q1 b q2) (q2 a q1) (q2 b q2)
                              (q3 a q4) (q3 b q4) (q4 a q4)
                              (q4 b q4) (ds a ds) (ds b ds))))

(define remove-unreachable-test1 (remove-unreachable dfa-test1))

(check-expect (sm-testequiv dfa-test1 remove-unreachable-test1) #t)
\end{alltt}
\caption{A student's transformer to remove unreachable states from a deterministic finite-state automaton.}
\label{us}
\end{figure}

The final example involves removing all unreachable states from a deterministic finite-state automaton. An outline of a student's proposed solution is displayed in Figure \ref{us}. The student has successfully implemented a function to compute the unreachable states using breath-first search. The student also provides a sample deterministic finite-state automaton and uses \fsm to test if it is equivalent to the result obtained from removing the unreachable states. No errors are reported. There are, however, two type bugs. The constructor \textsf{make-dfa} is not properly used. Unreachable states are neither removed from the list of final states nor are rules involving unreachable states removed. These are dependent type errors that \fsm can and ought to handle better. Students should get an error when trying to build machines with rules involving nonexisting states. The same holds true for constructing machines with accepting states that are not in the list of states for the machine.

\section{Design of \fsm  Error Messages}
\label{fsmerrs}
The first version of the \fsm error-messaging system is designed around building guarded constructors for state machines and grammars. This decision follows from the errors made by students observed by the instructor during the maiden voyage of \fsm at \textsf{SHU}. The overwhelming majority of errors revolved around misuses of constructors as those outlined in the previous section. Students expressed a great deal of frustration by having no real way to understand the error messages on their own. This meant students had to see the instructor every week to complete homework assignments and take-home quizzes. Although having office hours packed with students may be the dream of a true educator, having students come mostly to help them debug misuses of \fsm constructors was nothing less than disappointing. Students, in fact, eventually learned how to fix errors by memorization rather than by the usefulness of the error messages. Students also realized they all faced, in essence, the same error messages and developed a system to take turns to attend office hours. One group would come to office hours and then this group would communicate the real meaning of an error message to the rest of the class.

It was clear that for the next use of \fsm it was necessary to develop meaningful error messages. It is noteworthy that all students were upper level undergraduates. This suggests that meaningful error messages are needed at all levels of the undergraduate curriculum. Some argue that compilers and libraries ought to always generate meaningful error messages \cite{Marceau,Mind,Traver}. With this in mind, we chose to endow \fsm error messages with the following characteristics:
\begin{description}
  \item[Vocabulary] The vocabulary of the error messages must match the vocabulary the instructor uses in the classroom. This means that the instructors must make an effort to use the same automata theory jargon when developing machines, grammars, and constructive proofs in class.
  \item[Length] The decision was made to have longer informative error messages. Instead of reporting errors piecemeal (i.e., one error at a time), all errors caused by the misuse of a constructor are reported at once. This is a feature that students on the maiden voyage felt they would find most useful. Furthermore, it makes sense, because the inputs to the constructor are dependent on each other.
  \item[Ordering] For state-machine constructors, the starting state, the list of final states, and the list of rules depend on the list of states, the alphabet, and the stack alphabet (for pushdown automata only). Errors for the non-dependent types are listed first. Similarly, for grammar constructors the errors pertaining to \textsf{V}, the set of nonterminals, and to \textsf{$\Sigma$}, the alphabet, are reported first. The expectation is that listing errors with the non-dependent types first may make determining bugs easier.
  \item[Prescription] The \fsm error messages should not prescribe solutions as suggested by Marceau et al. \cite{Mind}.
  \item[Highlighting] A misused constructor must be highlighted by \textsf{DrRacket} when an error is thrown. This means that errors must be thrown by the constructor and not by an auxiliary function used by the constructor.
\end{description}

The implementation of the error-messaging system is completely hidden from the user and is designed to operate on top of the existing \textsf{FSM} library. Constructor wrappers, that test the validity of the input received, serve as guards. If the input received is valid, then the unguarded constructor is called. If the input received is not valid, then the wrapper throws an error and prints a message. In this manner, improvements in the error-messaging system and in the \textsf{FSM} constructors may be developed in tandem.

For the Turing machine constructed in Figure \ref{tmaa}, the error messages reported now looks like this:
\begin{alltt}
     the alphabet sigma must be a list
     the given rule a must be a list
     the given rule b must be a list
\end{alltt}
These error messages are clearly communicating that the alphabet and the rules are not lists. This clues students into the fact that the wrong types of arguments are being provided as input to the constructor. It is noteworthy that \textsf{DrRacket} highlights the use of \textsf{make-tm} in the file students are running.

For the context-free grammar constructed in Figure \ref{pali}, the errors reported now look like this:
\begin{alltt}
     b must be uppercase to be valid for V
     s must be uppercase to be valid for V
     S must be a symbol in V to be a valid lhs for the rule (S -> Bsb)
     The symbols (B) are not part of V or sigma.
     S must be a symbol in V to be a valid lhs for the rule (S -> aSa)
     The symbols (S) are not part of V or sigma.
     S must be a symbol in V to be a valid lhs for the rule (S -> EMP)
     The symbols (E M P) are not part of V or sigma.
     (S) must be in V to be a valid starting symbol
\end{alltt}
The first two messages indicate that elements in \textsf{V}, the set of nonterminals, are not properly capitalized. The next 6 messages are indicating errors in the list of rules stating that there are elements in the rules that are neither in the set of nonterminal nor in sigma (the alphabet of terminal symbols). For a given rule, left-hand side (\textsf{lhs}) errors are listed first followed by errors from the right-hand side (\textsf{rhs}). The eighth message may be of particular interest. It is explained to students that the right-hand side of a rule is decomposed (i.e., parsed) into \fsm symbols. The error is indicating that \textsf{E}, \textsf{M}, and \textsf{P} are not recognized as valid symbols in the grammar. Notice that it is not suggesting that the student may have not used a \textsf{quasiquote} to recognize the \fsm constant \textsf{EMP}. Such a suggestion would be improper, because it is equally feasible, for example, that the student simply forgot to include \textsf{E}, \textsf{M}, and \textsf{P} in the list of nonterminal symbols. The final message indicates that the argument provided as the starting nonterminal is not in the in the list of nonterminal symbols. For this final error message, it would be possible to report that he argument must be a symbol. The decision was made to use the displayed format to keep individual messages short. None of the error messages, as the reader may expect, address the fact that the constructed grammar does not generate the correct language.

For the finite-state automaton constructed in Figure \ref{us}, the error messages are not empty. The reported errors are as follows:
\begin{alltt}
     the state q3 in the rule (q3 a q4) must be in your list of states
     the state q4 in the rule (q3 a q4) must be in your list of states
     the state q3 in the rule (q3 b q4) must be in your list of states
     the state q4 in the rule (q3 b q4) must be in your list of states
     the state q4 in the rule (q4 a q4) must be in your list of states
     the state q4 in the rule (q4 a q4) must be in your list of states
     the state q4 in the rule (q4 b q4) must be in your list of states
     the state q4 in the rule (q4 b q4) must be in your list of states
     (q3) cannot be final states because they are not in the initial
        list of states for the machine
\end{alltt}
\textsf{DrRacket} highlights the error as coming from the use of the constructor \textsf{make-dfa} in the body of \textsf{remove-unreachable}. The first eight messages indicate that states \textsf{q3} and \textsf{q4}, which are not states in the machine, are being used in the rules of the machine. No corrective action is suggested. Students must decide if these states need to be or not be in the list of states for the machine. In this case, these are the unreachable states and this should help students make progress towards correctly solving the problem allowing them to realize that their transformer must remove rules involving unreachable states. The final message is indicating that nonexisting states are listed as final states. Once again, no corrective action is suggested. The error should help students realize that unreachable states must also be removed from the list of final states.

\pgfplotstableread[row sep=\\,col sep=&]{
size    & R-ERRS		& FSM-ERRS	   \\
TMBB	& 0.8		    & 0.67			\\
PALI	& 1.7		    & 1.56			\\
RMUN	& 2		        & 1.56			\\
}\ctrlstudy

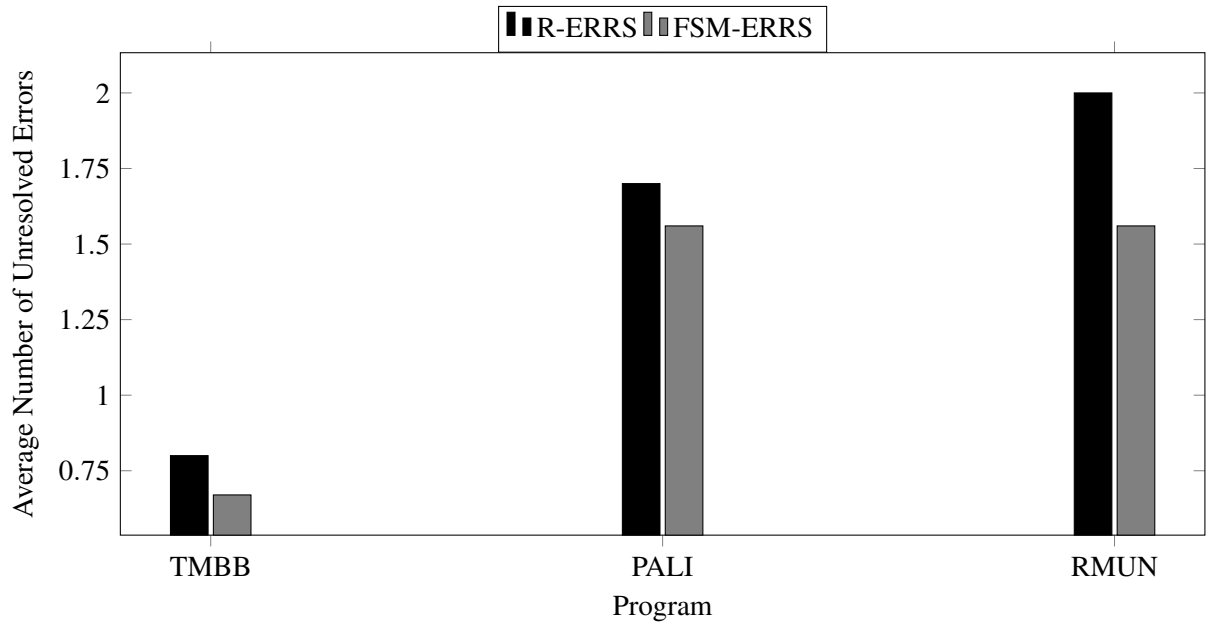
\begin{figure}[t]
\begin{tikzpicture}

    \begin{axis}[
            ybar,
            bar width=.5cm,
            width=\textwidth,
            height=.5\textwidth,
            legend style={at={(0.5,1)},
               anchor=south,legend columns=-1},
            symbolic x coords={0,TMBB,PALI,RMUN},
            xtick=data,
            ytick={0,0.25,0.50,...,2.0},
            nodes near coords align={vertical},
            ylabel={Average Number of Unresolved Errors},
            xlabel={Program},
        ]
        \addplot[fill=black] table[x=size,y=R-ERRS]{\ctrlstudy};
        \addplot[fill=black!50] table[x=size,y=FSM-ERRS]{\ctrlstudy};
        \legend{R-ERRS, FSM-ERRS}
    \end{axis}
  \end{tikzpicture}
\caption{Average Number of Unresolved Errors After Debugging.} \label{ctrl}
\end{figure}

\section{Student Feedback}
\label{fb}
Students feedback was obtained in two ways. The first method was a control group study to determine the effectiveness of \fsm error messages. Students in the second voyage of \fsm at \textsf{SHU} in the Spring 2018 semester were evenly divided at the end of the semester into two groups. Each group was asked to debug the three examples described in Section \ref{pe}. One group was assigned to work with the version of \fsm that produces \racket error messages (described in Section \ref{pe}). The other group was assigned to work with the version that produces \fsm error messages (described in Section \ref{fsmerrs}). Of the 25 students enrolled 19 volunteered to participate in the study of which 10 were randomly placed in the group that used \fsm with \racket error messages and 9 in the group that used the library that produces \fsm error messages.

The second method aims to ascertain student attitudes towards the course, towards programming with \fsmn, and towards \fsm error messages. Students of both the maiden (\textsf{V1}) and the second voyage (\textsf{V2}) of \fsm at \textsf{SHU} were surveyed. For the maiden voyage, 100\% of the students (8/8) chose to participate in the survey. For the second voyage, 92\% of the students (23/25) chose to participate in the survey.

\subsection{Control Group Study}
\label{cg}

Figure \ref{ctrl} displays the average number of unresolved errors for each program after debugging. The x-axis represents the programs students attempted to debug. The programs are labeled as follows:
\begin{description}
  \item[TMBB] The Turing machine that decides the language of all strings that have \textsf{bb} as a substring.
  \item[PALI] The context-free grammar for the language of palindromes.
  \item[RMUN] The function to remove unreachable states from a deterministic finite-state automaton.
\end{description}
The y-axis is the average number of unresolved errors for each program after debugging. For each program there are two columns. The first is for the control group using \fsm with \racket error messages (\textsf{R-ERRS}). The second is for the study group using the library with \fsm error messages (\textsf{FSM-ERRS}). The students were given one class period (i.e., 75 minutes) to debug the three functions. They were advised to spend at most 25 minutes on each program.

For \textsf{TMBB}, the average number of unresolved errors after debugging is less than 1 for both groups. That is, most students in both groups fully debugged the Turing machine implementation. The majority of students in the study group commented that the \fsm error messages were useful in honing in on the problems. In the control group, virtually all students commented that the error messages were ``useless." These students also commented that the only reason they were able to debug the function was due to their prior experience using \textsf{make-tm}. In essence, they remembered or looked up in the documentation the correct order for arguments. The study group, nonetheless, exhibits 20\% fewer errors left unresolved when compared to the control group. This indicates that \fsm error messages have a significant impact on helping students debug programs.

\pgfplotstableread[row sep=\\,col sep=&]{
size            & V1		& V2	   \\
1	            & 0.125		& 0.00			\\
2	            & 0.125		& 0.04			\\
3	            & 0.25		& 0.22			\\
4	            & 0.50		& 0.61			\\
5	            & 0.00		& 0.13			\\
}\cop

\begin{figure}[t]
\begin{tikzpicture}

    \begin{axis}[
            ybar,
            bar width=.5cm,
            width=\textwidth,
            height=.5\textwidth,
            legend style={at={(0.5,1)},
               anchor=south,legend columns=-1},
            symbolic x coords={0,1,2,3,4,5},
            xtick=data,
            ytick={0,0.1,0.2,...,1.0},
            nodes near coords align={vertical},
            ylabel={Proportion of Students},
            xlabel={Opinion},
        ]
        \addplot[fill=black] table[x=size,y=V1]{\cop};
        \addplot[fill=black!50] table[x=size,y=V2]{\cop};
        \legend{V1,V2}
    \end{axis}
  \end{tikzpicture}
\caption{Overall Course Opinion.} \label{co}
\end{figure}
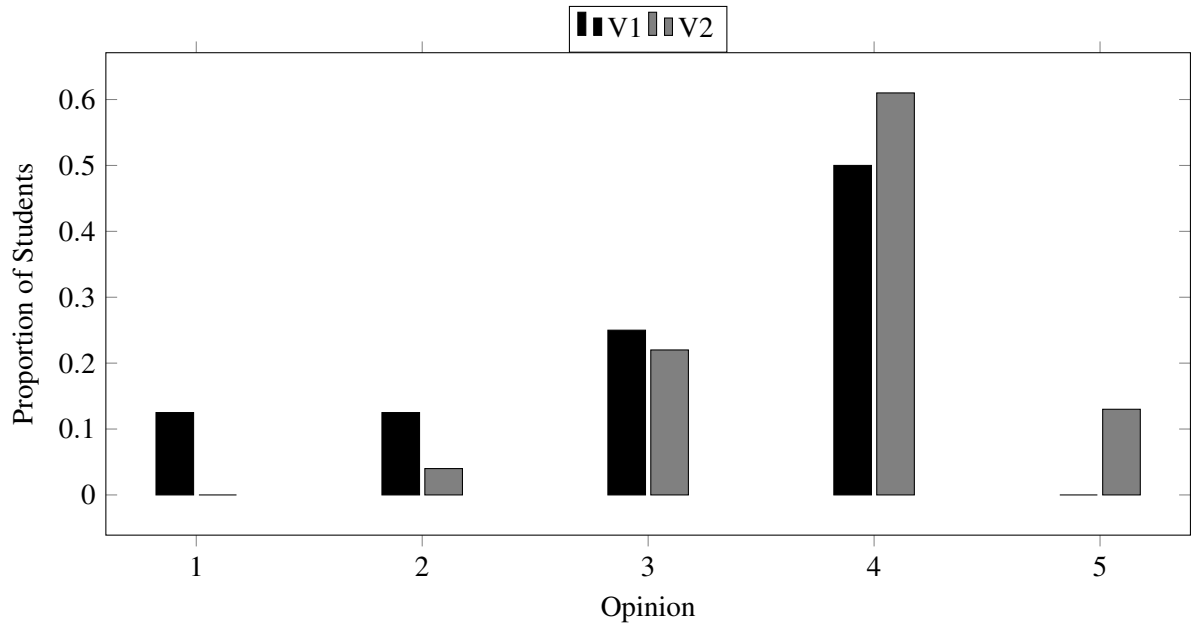

For \textsf{PALI}, both groups had, on average, resolved 2-3 of the 4 bugs. The bug that was most frequently left outstanding in both groups was fixing the language that the grammar generated (i.e., to also generate palindromes of odd length). This is probably not a big surprise as there are no \fsm or \racket error messages generated for this bug. The next most frequently unresolved error was the correct capitalization of nonterminals. In both groups, the majority correctly capitalized the starting nonterminal to \textsf{S} in the list of nonterminals and in the rules. In the control group, most students did not change \textsf{B} to \textsf{b} in the rules nor did they remove \textsf{b} from the list of nonterminals. Something interesting happened in the study group. After correctly capitalizing \textsf{S}, the \fsm error messages they receive are:
\begin{alltt}
     b must be uppercase to be valid for V
     The symbols (B) are not part of V or sigma.
     The symbols (E M P) are not part of V or sigma.
\end{alltt}
The third message was usually the first students tackled. The second message refers to bugs in the rules. The corrective action many students took was to capitalize \textsf{b} in the list of nonterminals. After this corrective action, they did not receive any error messages (if they had already resolved the \textsf{EMP} error) or they received:
\begin{alltt}
     The symbols (E M P) are not part of V or sigma.
\end{alltt}
Regardless of the order in which reported errors were resolved, most students considered the grammar successfully debugged. Only a handful of students observed that something is still wrong, because they unsuccessfully tried to derive palindromes using \fsmn's \textsf{grammar-derive}. These students complained that no error messages were reported for palindromes of odd length. This indicates that students are focusing on getting fewer error messages and not on trying to understand if there is a design bug in their grammar. It suggests that instructors would be well-advised to constantly remind students that it does not suffice to only eliminate type errors. Despite these issues, the study group exhibits 9\% fewer errors left unresolved when compared to the control group. Once again, this indicates that \fsm error messages have a positive impact on helping students debug programs.

For \textsf{RMUN}, students were told that they may assume that the list of unreachable states is correctly computed by the function \textsf{reachable-states}. For this program, we observe the largest gap between the two groups with the study group having on average 29\% fewer outstanding bugs. All students in the control group and the majority in the study group, however, did not resolve either of the two bugs. In the control group, only 2 students realized that the rules needed to be filtered and none realized that the set of final states also needed to be filtered to eliminate unreachable states. Eight students stated that they received no error messages and that the machine worked correctly. This indicates that students, do indeed, need informative error messages to find implementation shortcomings. In the study group all but 1 student knew what needed to be done to eliminate the bugs. The majority, however, did not get around to implementing the corrective actions. This indicates that the \fsm error messages are effective in finding implementation shortcomings. These students simply needed more time to implement the solution.

\subsection{Student Survey}
\label{ss}

Students were asked what is their overall evaluation of Automata Theory course on a scale from \emph{1 = A complete waste of time} to \emph{5 = Beyond Excellent}. The results are displayed in Figure \ref{co}. Half, 50\%, of the students on the maiden voyage of \fsm at \textsf{SHU} highly ranked the course (responses in 4-5). In contrast, the overwhelming majority of students, 74\%, on the second voyage highly ranked the course. The only significant differences in the delivery of both versions of the course is the use of the tailored-made error-messaging system and the associated coverage of error messages in class using the error-message vocabulary. In both groups, qualitative responses stated that the best part of the course was the programming component. The students on the first voyage, however, did complain about how often they had to go to office hours. All this indicates that \fsm error messages have a positive impact on how receptive students are to the course.

\pgfplotstableread[row sep=\\,col sep=&]{
size            & V1		& V2	   \\
1	            & 0.00		& 0.04			\\
2	            & 0.375		& 0.09			\\
3	            & 0.375		& 0.22			\\
4	            & 0.125		& 0.43			\\
5	            & 0.125		& 0.22		\\
}\fsmop

\begin{figure}[t]
\begin{tikzpicture}

    \begin{axis}[
            ybar,
            bar width=.5cm,
            width=\textwidth,
            height=.5\textwidth,
            legend style={at={(0.5,1)},
               anchor=south,legend columns=-1},
            symbolic x coords={0,1,2,3,4,5},
            xtick=data,
            ytick={0,0.1,0.2,...,1.0},
            nodes near coords align={vertical},
            ylabel={Proportion of Students},
            xlabel={Interest},
        ]
        \addplot[fill=black] table[x=size,y=V1]{\fsmop};
        \addplot[fill=black!50] table[x=size,y=V2]{\fsmop};
        \legend{V1,V2}
    \end{axis}
  \end{tikzpicture}
\caption{Level of Interest in Programming with \fsm.} \label{opfsm}
\end{figure}
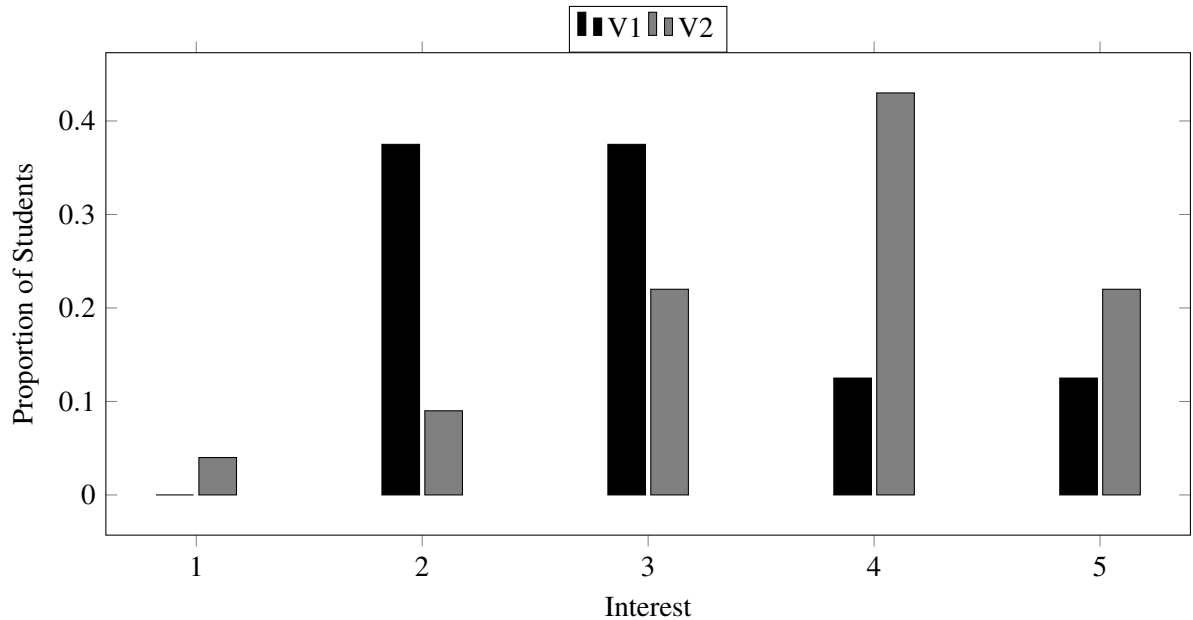

Students were asked to overall evaluate their level of interest in programming with \fsm on a scale from \emph{1 = Not at all interested} to \emph{5 = Very interested}. The results are displayed in Figure \ref{opfsm}. In the maiden voyage at \textsf{SHU}, only a minority of students, 25\%, signaled a significant interest in programming with \fsm (responses in 4-5). The overwhelming majority of students, 75\%, had, at best, a lukewarm interest in programming with \fsm (responses 2-3). In contrast, for the second delivery of the course using \fsm error messages an overwhelming majority of students, 65\%, signaled a significant interest in programming with \fsm and a minority, 31\%, signaled a lukewarm interest. This drastic change in observed attitudes can only be attributed to having a better error-messaging system. It is important to note that students are specifically expressing an interest in programming with \fsmn. The lesson that needs to be derived is that even libraries for advanced courses benefit from a well-designed error-messaging system.

\begin{figure}
\centering
\begin{tikzpicture}
\selectcolormodel{gray}
\pie [rotate = 180]
    {
     4/2,
     39/3,
     35/4,
     22/5}
\end{tikzpicture}
\caption{Overall  Opinion on \fsm Error Messages.} \label{fsmerrop}
\end{figure}
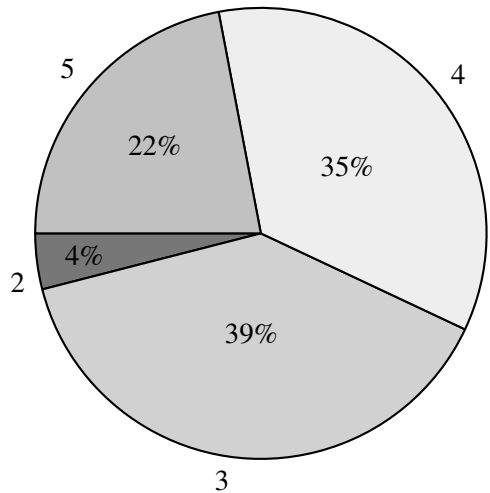

Students in the second voyage of \fsm at \textsf{SHU} were asked to evaluate their overall opinion of \fsm error messages on a scale from \emph{1 = A complete waste of time} to \emph{5 = Beyond excellent}\footnote{Students in the maiden voyage of \fsm were not asked this question, because error messages were not on our radar.}. The results are displayed in Figure \ref{fsmerrop}. The responses clearly suggest that students like the \fsm error messages. A majority of students, 57\%, signaled a very positive opinion (responses in 4-5). In fact, 96\% of students signaled a positive opinion (responses 3-5). This result, in our opinion, is the most important reason that explains why students have a more positive attitude towards Automata Theory and programming in \fsmn. We also have no doubt that having the instructor cover error messages and use error-message vocabulary in class helped make the error-messaging system a success.

\section{Concluding Remarks and Future Work}
\label{fw}
This article presents the design of \fsm error messages. These messages aim to help students diagnose the reason for bugs without prescribing a solution. The error messages revolve around type violations when using an \fsm constructor. The system reports all errors that are found instead of just the first one. It also specifically avoids reporting parsing errors and list errors for non-dependent types before errors for dependent types. The empirical data collected strongly suggests that the \fsm error-messaging system is very successful. It has successfully contributed to increasing students' interest in Automata Theory and in programming in \fsmn. Furthermore, almost all students report a positive opinion about \fsm error messages. The work strongly suggests that an error-messaging system tailored for students is as important in advanced courses as it is in courses for novices.

Future work includes exploring how to improve the wording of individual messages. For example, should lots of detail be provided or should less detail be provided? Currently, \fsm error messages provide lots of detail when machine rules contain type errors, but much less detail is provided for grammar rules. More qualitative research is needed to determine if one strategy is better than the other. To date, we can see no difference. That, however, may be due to the instructor covering error messages in his lectures. Future work also includes collecting more quantitative data to strengthen or modify the conclusions suggested in this article.

\section{Acknowledgements}
The authors thank all the Automata Theory students that accepted to participate in this study. We also express our gratitude to the Department of Computer Science and to the Office of the Dean of The College of Arts and Sciences of Seton Hall University that made this work and the travel to present results possible.

\bibliographystyle{eptcs}
\bibliography{fsmerrors}
\end{document}